\documentstyle[aps,preprint]{revtex}

\begin{document}
\author{Remo Garattini}
\address{M\'{e}canique et Gravitation, Universit\'{e} de Mons-Hainaut,\\
Facult\'e des Sciences, 15 Avenue Maistriau, \\
B-7000 Mons, Belgium \\
and\\
Facolt\`a di Ingegneria, Universit\`a degli Studi di Bergamo,\\
Viale Marconi, 5, 24044 Dalmine (Bergamo) Italy\\
e-mail: Garattini@mi.infn.it}
\title{Large $N$-wormhole approach to spacetime foam}
\date{\today}
\maketitle

\begin{abstract}
A simple model of spacetime foam, made by $N$ wormholes in a semiclassical
approximation, is taken under examination. We show that the qualitative
behaviour of the fluctuation of the metric conjectured by Wheeler is here
reproduced.
\end{abstract}

\section{Introduction}

One of the most fascinating problem of our century is the possibility of
combining the principles of Quantum Mechanics with those of General
Relativity. The result of this combination is best known as Quantum Gravity.
However such a theory has to be yet developed, principally due to the UV
divergences that cannot be kept under control by any renormalization scheme.
J.A. Wheeler\cite{Wheeler} was the first who conjectured that fluctuations
of the metric have to appear at short scale distances. The collection of
such fluctuations gives the spacetime a kind of foam-like structure, whose
topology is constantly changing. In this foamy spacetime a fundamental
length comes into play: the Planck length. Its inverse, the Planck mass $m_p$%
, can be thought as a natural cut-off. It is believed that in such
spacetime, general relativity can be renormalized when a density of virtual
black holes is taken under consideration coupled to $N$ fermion fields in a $%
1/N$ expansion\cite{CraneSmolin}. It is also argued that when gravity is
coupled to $N$ conformally invariant scalar fields the evidence that the
ground-state expectation value of the metric is flat space is false\cite
{HartleHorowitz}. However instead of looking at gravity coupled to matter
fields, we will consider pure gravity. In this context two metrics which are
solutions of the equations of motion without a cosmological constant are
known with the property of the spherical symmetry: the Schwarzschild metric
and the Flat metric. We will focus our attention on these two metrics with
the purpose of examining the energy contribution to the vacuum fluctuation
generated by a collection of $N$ coherent wormholes. A straightforward
extension to the deSitter and the Schwarzschild-deSitter spacetime case is
immediate. The paper is structured as follows, in section \ref{p2} we
briefly recall the results reported in Ref.\cite{Remo1}, in section \ref{p3}
we generalize the result of section \ref{p2} to $N_w$ wormholes. We
summarize and conclude in section \ref{p3}.

\section{One wormhole approximation}

\label{p2}The reference model we will consider is an eternal black hole. The
complete manifold ${\cal M}$ can be thought as composed of two wedges ${\cal %
M}_{+}$ and ${\cal M}_{-}$ located in the right and left sectors of a
Kruskal diagram whose spatial slices $\Sigma $ represent Einstein-Rosen
bridges with wormhole topology $S^2\times R^1$. The hypersurface $\Sigma $
is divided in two parts $\Sigma _{+}$ and $\Sigma _{-}$ by a bifurcation
two-surface $S_0$. We begin with the line element 
\begin{equation}
ds^2=-N^2\left( r\right) dt^2+\frac{dr^2}{1-\frac{2m}r}+r^2\left( d\theta
^2+\sin ^2\theta d\phi ^2\right)  \label{a1}
\end{equation}
and we consider the physical Hamiltonian defined on $\Sigma $%
\[
H_P=H-H_0=\frac 1{l_p^2}\int_\Sigma d^3x\left( N{\cal H}+N_i{\cal H}%
^i\right) +H_{\partial \Sigma ^{+}}+H_{\partial \Sigma ^{-}} 
\]
\[
=\frac 1{l_p^2}\int_\Sigma d^3x\left( N{\cal H}+N_i{\cal H}^i\right) 
\]
\begin{equation}
+\frac 2{l_p^2}\int_{S_{+}}^{}d^2x\sqrt{\sigma }\left( k-k^0\right) -\frac 2{%
l_p^2}\int_{S_{-}}d^2x\sqrt{\sigma }\left( k-k^0\right) ,
\end{equation}
where $l_p^2=16\pi G$. The volume term contains two contstraints 
\begin{equation}
\left\{ 
\begin{array}{l}
{\cal H}=G_{ijkl}\pi ^{ij}\pi ^{kl}\left( \frac{l_p^2}{\sqrt{g}}\right)
-\left( \frac{\sqrt{g}}{l_p^2}\right) R^{\left( 3\right) }=0 \\ 
{\cal H}^i=-2\pi _{|j}^{ij}=0
\end{array}
\right. ,  \label{a1a}
\end{equation}
where $G_{ijkl}=\frac 12\left( g_{ik}g_{jl}+g_{il}g_{jk}-g_{ij}g_{kl}\right) 
$ and $R^{\left( 3\right) }$ denotes the scalar curvature of the surface $%
\Sigma $. By using the expression of the trace 
\begin{equation}
k=-\frac 1{\sqrt{h}}\left( \sqrt{h}n^\mu \right) _{,\mu },
\end{equation}
with the normal to the boundaries defined continuously along $\Sigma $ as $%
n^\mu =\left( h^{yy}\right) ^{\frac 12}\delta _y^\mu $. The value of $k$
depends on the function $r,_y$, where we have assumed that the function $%
r,_y $ is positive for $S_{+}$ and negative for $S_{-}$. We obtain at either
boundary that 
\begin{equation}
k=\frac{-2r,_y}r.
\end{equation}
The trace associated with the subtraction term is taken to be $k^0=-2/r$ for 
$B_{+}$ and $k^0=2/r$ for $B_{-}$. Then the quasilocal energy with
subtraction terms included is 
\begin{equation}
E_{{\rm quasilocal}}=E_{+}-E_{-}=\left( r\left[ 1-\left| r,_y\right| \right]
\right) _{y=y_{+}}-\left( r\left[ 1-\left| r,_y\right| \right] \right)
_{y=y_{-}}.
\end{equation}
Note that the total quasilocal energy is zero for boundary conditions
symmetric with respect to the bifurcation surface $S_0$ and this is the
necessary condition to obtain instability with respect to the flat space. A
little comment on the total Hamiltonian is useful to further proceed. We are
looking at the sector of asymptotically flat metrics included in the space
of all metrics, where the Wheeler-DeWitt equation 
\begin{equation}
{\cal H}\Psi =0
\end{equation}
is defined. In this sector the Schwarzschild metric and the Flat metric
satisfy the constraint equations $\left( \ref{a1a}\right) $. Here we
consider deviations from such metrics in a WKB approximation and we
calculate the expectation value following a variational approach where the
WKB functions are substituted with trial wave functionals. Then the
Hamiltonian referred to the line element $\left( \ref{a1}\right) $ is 
\[
H=\int_\Sigma d^3x\left[ G_{ijkl}\pi ^{ij}\pi ^{kl}\left( \frac{l_p^2}{\sqrt{%
g}}\right) -\left( \frac{\sqrt{g}}{l_p^2}\right) R^{\left( 3\right) }\right]
. 
\]
Instead of looking at perturbations on the whole manifold ${\cal M}$, we
consider perturbations at $\Sigma $ of the type $g_{ij}=\bar{g}_{ij}+h_{ij}$%
. $\bar{g}_{ij}$ is the spatial part of the background considered in eq.$%
\left( \ref{a1}\right) $In Ref.\cite{Remo1}, we have defined $\Delta E\left(
m\right) $ as the difference of the expectation value of the Hamiltonian
approximated to second order calculated with respect to different
backgrounds which have the asymptotic flatness property. This quantity is
the natural extension to the volume term of the subtraction procedure for
boundary terms and is interpreted as the Casimir energy related to vacuum
fluctuations. Thus 
\[
\Delta E\left( m\right) =E\left( m\right) -E\left( 0\right) 
\]
\begin{equation}
=\frac{\left\langle \Psi \left| H^{Schw.}-H^{Flat}\right| \Psi \right\rangle 
}{\left\langle \Psi |\Psi \right\rangle }+\frac{\left\langle \Psi \left|
H_{quasilocal}\right| \Psi \right\rangle }{\left\langle \Psi |\Psi
\right\rangle }.
\end{equation}
By restricting our attention to the graviton sector of the Hamiltonian
approximated to second order, hereafter referred as $H_{|2}$, we define 
\[
E_{|2}=\frac{\left\langle \Psi ^{\perp }\left| H_{|2}^1\right| \Psi ^{\perp
}\right\rangle }{\left\langle \Psi ^{\perp }|\Psi ^{\perp }\right\rangle }, 
\]
where 
\[
\Psi ^{\perp }=\Psi \left[ h_{ij}^{\perp }\right] ={\cal N}\exp \left\{ -%
\frac 1{4l_p^2}\left[ \left\langle \left( g-\bar{g}\right) K^{-1}\left( g-%
\bar{g}\right) \right\rangle _{x,y}^{\perp }\right] \right\} . 
\]
After having functionally integrated $H_{|2}$, we get 
\begin{equation}
H_{|2}=\frac 1{4l_p^2}\int_\Sigma d^3x\sqrt{g}G^{ijkl}\left[ K^{-1\bot
}\left( x,x\right) _{ijkl}+\left( \triangle _2\right) _j^aK^{\bot }\left(
x,x\right) _{iakl}\right]
\end{equation}
The propagator $K^{\bot }\left( x,x\right) _{iakl}$ comes from a functional
integration and it can be represented as 
\begin{equation}
K^{\bot }\left( \overrightarrow{x},\overrightarrow{y}\right) _{iakl}:=\sum_N%
\frac{h_{ia}^{\bot }\left( \overrightarrow{x}\right) h_{kl}^{\bot }\left( 
\overrightarrow{y}\right) }{2\lambda _N\left( p\right) },
\end{equation}
where $h_{ia}^{\bot }\left( \overrightarrow{x}\right) $ are the
eigenfunctions of 
\begin{equation}
\left( \triangle _2\right) _j^a:=-\triangle \delta _j^{a_{}^{}}+2R_j^a.
\end{equation}
This is the Lichnerowicz operator projected on $\Sigma $ acting on traceless
transverse quantum fluctuations and $\lambda _N\left( p\right) $ are
infinite variational parameters. $\triangle $ is the curved Laplacian
(Laplace-Beltrami operator) on a Schwarzschild background and $R_{j\text{ }%
}^a$ is the mixed Ricci tensor whose components are:

\begin{equation}
R_j^a=diag\left\{ \frac{-2m}{r_{}^3},\frac m{r_{}^3},\frac m{r_{}^3}\right\}
.
\end{equation}
After normalization in spin space and after a rescaling of the fields in
such a way as to absorb $l_p^2$, $E_{|2}$ becomes in momentum space 
\begin{equation}
E_{|2}\left( m,\lambda \right) =\frac V{2\pi ^2}\sum_{l=0}^\infty
\sum_{i=1}^2\int_0^\infty dpp^2\left[ \lambda _i\left( p\right) +\frac{%
E_i^2\left( p,m,l\right) }{\lambda _i\left( p\right) }\right] ,  \label{a3}
\end{equation}
where 
\begin{equation}
E_{1,2}^2\left( p,m,l\right) =p^2+\frac{l\left( l+1\right) }{r_0^2}\mp \frac{%
3m}{r_0^3}
\end{equation}
and $V$ is the volume of the system. $r_0$ is related to the minimum radius
compatible with the wormhole throat. We know that the classical minimum is
achieved when $r_0=2m$. However, it is likely that quantum processes come
into play at short distances, where the wormhole throat is defined,
introducing a {\it quantum} radius $r_0>2m$. The minimization with respect
to $\lambda $ leads to $\bar{\lambda}_i\left( p,l,m\right) =\sqrt{%
E_i^2\left( p,m,l\right) }$ and eq.$\left( \ref{a3}\right) $ becomes 
\begin{equation}
E_{|2}\left( m,\lambda \right) =2\frac V{2\pi ^2}\sum_{l=0}^\infty
\sum_{i=1}^2\int_0^\infty dpp^2\sqrt{E_i^2\left( p,m,l\right) },
\end{equation}
with $p^2+\frac{l\left( l+1\right) }{r_0^2}>\frac{3m}{r_0^3}.$ Thus, in
presence of the curved background, we get 
\begin{equation}
E_{|2}\left( m\right) =\frac V{2\pi ^2}\frac 12\sum_{l=0}^\infty
\int_0^\infty dpp^2\left( \sqrt{p^2+c_{-}^2}+\sqrt{p^2+c_{+}^2}\right)
\end{equation}
where 
\[
c_{\mp }^2=\frac{l\left( l+1\right) }{r_0^2}\mp \frac{3m}{r_0^3}, 
\]
while when we refer to the flat space, we have $m=0$ and $c^2=$ $\frac{%
l\left( l+1\right) }{r_0^2}$, with 
\begin{equation}
E_{|2}\left( 0\right) =\frac V{2\pi ^2}\frac 12\sum_{l=0}^\infty
\int_0^\infty dpp^2\left( 2\sqrt{p^2+c^2}\right) .
\end{equation}
Since we are interested in the $UV$ limit, we will use a cut-off $\Lambda $
to keep under control the $UV$ divergence 
\begin{equation}
\int_0^\infty \frac{dp}p\sim \int_0^{\frac \Lambda c}\frac{dx}x\sim \ln
\left( \frac \Lambda c\right) ,
\end{equation}
where $\Lambda \leq m_p.$ Note that in this context the introduction of a
cut-off at the Planck scale is quite natural if we look at a spacetime foam.
Thus $\Delta E\left( m\right) $ for high momenta becomes 
\begin{equation}
\Delta E\left( m\right) \sim -\frac V{2\pi ^2}\left( \frac{3m}{r_0^3}\right)
^2\frac 1{16}\ln \left( \frac{r_0^3\Lambda ^2}{3m}\right) .  \label{a4}
\end{equation}
We now compute the minimum of $\widetilde{\Delta E}\left( m\right) =E\left(
0\right) -E\left( m\right) =-\Delta E\left( m\right) $. We obtain two values
for $m$: $m_1=0$, i.e. flat space and $m_2=\Lambda ^2e^{-\frac 12}r_0^3/3$.
Thus the minimum of $\widetilde{\Delta E}\left( m\right) $ is at the value $%
\widetilde{\Delta E}\left( m_2\right) =\frac V{64\pi ^2}\frac{\Lambda ^4}e$.
Recall that $m=MG$, thus 
\begin{equation}
M=G^{-1}\Lambda ^2e^{-\frac 12}r_0^3/3.
\end{equation}
When $\Lambda \rightarrow m_p$, then $r_0\rightarrow l_p.$ This means that
an Heisenberg uncertainty relation of the type $l_pm_p=1$ (in natural units)
has to be satisfied, then 
\begin{equation}
M=m_p^2e^{-\frac 12}m_p^{-1}/3=\frac{m_p}{3\sqrt{e}}.
\end{equation}

\section{N$_{w}$ wormholes approximation}

\label{p3}

Suppose to consider $N_{w}$ wormholes and assume that there exists a
covering of $\Sigma $ such that $\Sigma =\cup _{i=1}^{N_{w}}\Sigma _{i}$,
with $\Sigma _{i}\cap \Sigma _{j}=\emptyset $ when $i\neq j$. Each $\Sigma
_{i}$ has the topology $S^{2}\times R^{1}$ with boundaries $\partial \Sigma
_{i}^{\pm }$ with respect to each bifurcation surface. On each surface $%
\Sigma _{i}$, quasilocal energy gives 
\begin{equation}
E_{i\text{ }{\rm quasilocal}}=\frac{2}{l_{p}^{2}}\int_{S_{i+}}d^{2}x\sqrt{%
\sigma }\left( k-k^{0}\right) -\frac{2}{l_{p}^{2}}\int_{S_{i-}}d^{2}x\sqrt{%
\sigma }\left( k-k^{0}\right) ,
\end{equation}
and by using the expression of the trace 
\begin{equation}
k=-\frac{1}{\sqrt{h}}\left( \sqrt{h}n^{\mu }\right) _{,\mu },
\end{equation}
we obtain at either boundary that 
\begin{equation}
k=\frac{-2r,_{y}}{r},
\end{equation}
where we have assumed that the function $r,_{y}$ is positive for $S_{i+}$
and negative for $S_{i-}$. The trace associated with the subtraction term is
taken to be $k^{0}=-2/r$ for $B_{i+}$ and $k^{0}=2/r$ for $B_{i-}$. Here the
quasilocal energy with subtraction terms included is 
\begin{equation}
E_{i\text{ }{\rm quasilocal}}=E_{i+}-E_{i-}=\left( r\left[ 1-\left|
r,_{y}\right| \right] \right) _{y=y_{i+}}-\left( r\left[ 1-\left|
r,_{y}\right| \right] \right) _{y=y_{i-}}.
\end{equation}
Note that the total quasilocal energy is zero for boundary conditions
symmetric with respect to {\it each} bifurcation surface $S_{0,i}$. We are
interested to a large number of wormholes, each of them contributing with a
Hamiltonian of the type $H_{|2}$. If the wormholes number is $N_{w}$, we
obtain (semiclassically, i.e., without self-interactions) 
\begin{equation}
H_{tot}^{N_{w}}=\underbrace{H^{1}+H^{2}+\ldots +H^{N_{w}}}.
\end{equation}
Thus the total energy for the collection is 
\[
E_{|2}^{tot}=N_{w}H_{|2}. 
\]
The same happens for the trial wave functional which is the product of $%
N_{w} $ t.w.f.. Thus 
\[
\Psi _{tot}^{\perp }=\Psi _{1}^{\perp }\otimes \Psi _{2}^{\perp }\otimes
\ldots \ldots \Psi _{N_{w}}^{\perp }={\cal N}\exp N_{w}\left\{ -\frac{1}{%
4l_{p}^{2}}\left[ \left\langle \left( g-\bar{g}\right) K^{-1}\left( g-\bar{g}%
\right) \right\rangle _{x,y}^{\perp }\right] \right\} 
\]
\[
={\cal N}\exp \left\{ -\frac{1}{4}\left[ \left\langle \left( g-\bar{g}%
\right) K^{-1}\left( g-\bar{g}\right) \right\rangle _{x,y}^{\perp }\right]
\right\} , 
\]
where we have rescaled the fluctuations $h=g-\bar{g}$ in such a way to
absorb $N_{w}/l_{p}^{2}.$ Of course, if we want the trial wave functionals
be independent one from each other, boundaries $\partial \Sigma ^{\pm }$
have to be reduced with the enlarging of the wormholes number $N_{w}$,
otherwise overlapping terms could be produced. Thus, for $N_{w}$-wormholes,
we obtain 
\[
H^{tot}=N_{w}H=\int_{\Sigma }d^{3}x\left[ G_{ijkl}\pi ^{ij}\pi ^{kl}\left(
N_{w}\frac{l_{p}^{2}}{\sqrt{g}}\right) -\left( N_{w}\frac{\sqrt{g}}{l_{p}^{2}%
}\right) R^{\left( 3\right) }\right] 
\]
\[
=\int_{\Sigma }d^{3}x\left[ G_{ijkl}\pi ^{ij}\pi ^{kl}\left( \frac{%
l_{N_{w}}^{2}}{\sqrt{g}}\right) -\left( N_{w}^{2}\frac{\sqrt{g}}{%
l_{N_{w}}^{2}}\right) R^{\left( 3\right) }\right] , 
\]
where we have defined $l_{N_{w}}^{2}=l_{p}^{2}N_{w}$ with $l_{N_{w}}^{2}$
fixed and $N_{w}\rightarrow \infty .$ Thus, repeating the same steps of
section \ref{p2} for $N_{w}$ wormholes, we obtain 
\begin{equation}
\Delta E_{N_{w}}\left( m\right) \sim -N_{w}^{2}\frac{V}{2\pi ^{2}}\left( 
\frac{3m}{r_{0}^{3}}\right) ^{2}\frac{1}{16}\ln \left( \frac{%
r_{0}^{3}\Lambda ^{2}}{3m}\right) .
\end{equation}
Then at one loop the cooperative effects of wormholes behave as one {\it %
macroscopic single }field multiplied by $N_{w}^{2}$; this is the consequence
of the coherency assumption. We have just explored the consequences of this
result in Ref.\cite{Remo1}$.$ Indeed, coming back to the single wormhole
contribution we have seen that the black hole pair creation probability
mediated by a wormhole is energetically favored with respect to the
permanence of flat space provided we assume that the boundary conditions be
symmetric with respect to the bifurcation surface which is the throat of the
wormhole. In this approximation boundary terms give zero contribution and
the volume term is nonvanishing. As in the one-wormhole case, we now compute
the minimum of $\widetilde{\Delta E}_{N_{w}}\left( m\right) =\left( E\left(
0\right) -E\left( m\right) \right) _{N_{w}}=-\Delta E_{N_{w}}\left( m\right) 
$. The minimum is reached for $\bar{m}=\Lambda ^{2}e^{-\frac{1}{2}%
}r_{0}^{3}/3$. Thus the minimum is 
\begin{equation}
\widetilde{\Delta E}\left( \bar{m}\right) =N_{w}^{2}\frac{V}{64\pi ^{2}}%
\frac{\Lambda ^{4}}{e}.
\end{equation}
The main difference with the one wormhole case is that we have $N_{w}$
wormholes contributing with the same amount of energy. Since $%
m=MN_{w}G=Ml_{N_{w}}^{2}$, thus 
\begin{equation}
M=\left( l_{N_{w}}^{2}/N_{w}\right) ^{-1}\Lambda ^{2}e^{-\frac{1}{2}%
}r_{0}^{3}/3.
\end{equation}
When $\Lambda \rightarrow m_{p}$, then $r_{0}\rightarrow l_{p}$ and $%
l_{p}m_{p}=1$. Thus 
\begin{equation}
M=\frac{\left( l_{N_{w}}^{2}/N_{w}\right) ^{-1}m_{p}^{-1}}{3\sqrt{e}}=N_{w}%
\frac{m_{N_{w}}}{3\sqrt{e}}
\end{equation}
So far, we have discussed the stable modes contribution. However, we have
discovered that for one wormhole also unstable modes contribute to the total
energy\cite{GPY,Remo1}. Since we are interested to a large number of
wormholes, the first question to answer is: what happens to the boundaries
when the wormhole number is enlarged. In the one wormhole case, the
existence of one negative mode is guaranteed by the vanishing of the
eigenfunction of the operator $\Delta _{2}$ at infinity, which is the same
space-like infinity of the quasilocal energy, i.e. we have the $ADM$
positive mass $M$ in a coordinate system of the universe where the observer
is present and the anti-$ADM$ mass in a coordinate system where the observer
is not there. When the number of wormholes grows, to keep the coherency
assumption valid, the space available for every single wormhole has to be
reduced to avoid overlapping of the wave functions. This means that boundary
conditions are not fixed at infinity, but at a certain finite radius and the 
$ADM$ mass term is substituted by the quasilocal energy expression under the
condition of having symmetry with respect to each bifurcation surface. As $%
N_{w}$ grows, the boundary radius $\bar{r}$ reduces more and more and the
unstable mode disappears. This means that there will exist a certain radius $%
r_{c}$ where below which no negative mode will appear and there will exist a
given value $N_{w_{c}}$ above which the same effect will be produced. In
rigorous terms: $\forall N\geq N_{w_{c}}\ \exists $ $r_{c}$ $s.t.$ $\forall
\ r_{0}\leq r\leq r_{c},\ \sigma \left( \Delta _{2}\right) =\emptyset $.
This means that the system begins to be stable. In support of this idea, we
invoke the results discovered in Ref. \cite{B.Allen} where, it is explicitly
shown that the restriction of spatial boundaries leads to a stabilization of
the system. Thus at the minimum, we obtain the typical energy density
behavior of the foam 
\begin{equation}
\frac{\Delta E}{V}\sim -N_{w}^{2}\Lambda ^{4}
\end{equation}

\section{Conclusions and Outlooks}

\label{p4}

According to Wheeler's ideas about quantum fluctuation of the metric at the
Planck scale, we have used a simple model made by a large collection of
wormholes to investigate the vacuum energy contribution needed to the
formation of a foamy spacetime. Such investigation has been made in a
semiclassical approximation where the wormholes are treated independently
one from each other (coherency hypothesis). The starting point is the single
wormhole, whose energy contribution has the typical trend of the
gravitational field energy fluctuation. The wormhole considered is of the
Schwarzschild type and every energy computation has to be done having in
mind the reference space, i.e. flat space. When we examine the wormhole
collection, we find the same trend in the energy of the single case. This is
obviously the result of the coherency assumption. However, the single
wormhole cannot be taken as a model for a spacetime foam, because it
exhibits one negative mode. This negative mode is the key of the topology
change from a space without holes (flat space) to a space with an hole
inside (Schwarzschild space). However things are different when we consider
a large number of wormholes $N_w$. Let us see what is going on: the
classical vacuum, represented by flat space is stable under nucleation of a
single black hole, while it is unstable under a neutral pair creation with
the components residing in \ different universes divided by a wormhole. When
the topology change has primed by means of a single wormhole, there will be
a considerable production of pairs mediated by their own wormhole. The
result is that the hole production will persist until the critical value $%
N_{w_c}$ will be reached and spacetime will enter the stable phase. If we
look at this scenario a little closer, we can see that it has the properties
of the Wheeler foam. Nevertheless, we have to explain why observations
measure a flat space structure. To this purpose, we have to recall that the
foamy spacetime structure should be visible only at the Planck scale, while
at greater scales it is likely that the flat structure could be recovered by
means of averages over the collective functional describing the {\it %
semiclassical} foam. Indeed if $\eta _{ij}$ is the spatial part of the flat
metric, ordinarily we should obtain 
\begin{equation}
\left\langle \Psi \left| g_{ij}\right| \Psi \right\rangle =\eta _{ij},
\end{equation}
where $g_{ij}$ is the spatial part of the gravitational field. However in
the foamy representation we should consider, instead of the previous
expectation value, the expectation value of the gravitational field
calculated on wave functional representing the foam, i.e., to see that at
large distances flat space is recovered we should obtain 
\begin{equation}
\left\langle \Psi _{foam}\left| g_{ij}\right| \Psi _{foam}\right\rangle
=\eta _{ij},
\end{equation}
where $\Psi _{foam}$ is a superposition of the single-wormhole wave
functional 
\begin{equation}
\Psi _{foam}=\sum_{i=1}^{N_w}\Psi _i^{\perp }.
\end{equation}
This has to be attributed to the semiclassical approximation which render
this system a non-interacting system. However, things can change when we
will consider higher order corrections and the other terms of the action
decomposition, i.e. the spin one and spin zero terms. Nevertheless, we can
argue that only spin zero terms (associated with the conformal factor) will
be relevant, even if the part of the action which carries the physical
quantities is that discussed in this text, i.e., the spin two part of the
action related to the gravitons.

\section{Acknowledgments}

I wish to thank R. Brout, M. Cavagli\`{a}, C. Kiefer, D. Hochberg, G.
Immirzi, S. Liberati, P. Spindel and M. Visser for useful comments and
discussions.

\end{document}